\documentclass{article}
\pdfoutput=1

\usepackage{arxiv}
\usepackage{graphicx}
\usepackage{amsmath}
\usepackage[utf8]{inputenc} 
\usepackage[T1]{fontenc}    
\usepackage{hyperref}       
\usepackage{url}            
\usepackage{booktabs}       
\usepackage{amsfonts}       
\usepackage{nicefrac}       
\usepackage{microtype}      

\title{Wavelength-multiplexed single-shot ptychography}

\author{
  Jonathan Barolak \\
  Department of Physics\\
  Colorado School of Mines\\
  Golden, CO 80401 \\
  \texttt{jjbarolak@mymail.mines.edu} \\
   \And
  David Goldberger\\
  Department of Physics\\
  Colorado School of Mines\\
  Golden, CO 80401 \\
   \And
  Jeff Squier \\
  Department of Physics\\
  Colorado School of Mines\\
  Golden, CO 80401 \\
   \And
  Yves Bellouard \\
  Galatea Laboratory\\
  Ecole Polytechnique Fédérale de Lausanne (EPFL)\\
  Neuchâtel, Switzerland \\
   \And
  Charles Durfee \\
  Department of Physics\\
  Colorado School of Mines\\
  Golden, CO 80401 \\
   \And
  Daniel Adams \\
  Department of Physics\\
  Colorado School of Mines\\
  Golden, CO 80401 \\
}

\begin{document}
\maketitle

\begin{abstract}
Diagnostics capable of interrogating dynamics in harsh environments such as plasma have remained essentially unchanged in recent decades. Developments in advanced microscopy techniques will improve our understanding of the physics involved in these events. Recently developed single-shot ptychography (SSP) provides a pathway towards sophisticated plasma metrologies. Here we introduce wavelength-multiplexed single-shot ptychography (WM-SSP), which allows for hyperspectral, spatially and temporally resolved phase and amplitude contrast imaging. Furthermore, we introduce a novel probe constraint common to all wavelength multiplexed modalities in the single-shot geometry and present modifications to SSP that improve reconstruction fidelity and robustness. WM-SSP was experimentally realized and simulations show the technique's ability to deconvolve the electron and neutral densities within the plasma. WM-SSP will pave the way to a new generation of quantitative plasma imaging techniques.
\end{abstract}

\section{Introduction}

Fundamental properties of dynamically evolving plasmas such as plasma channel size, electron and neutral density structure, and resistivity are challenging to measure \cite{Aragon2008}. Quantitative analysis of these basic characteristics is imperative to developing a robust theoretical framework of the dynamics involved in plasma formation and evolution. Common plasma imaging tools such as Schlieren imaging and interferometry are either not quantitative in nature, not directly sensitive to the phase, not spatially (in two dimensions) and temporally resolved, or restricted to weakly scattering plasmas \cite{Weber1997,Traldi2018,Biganzoli2015,Hayasaki2017,Aragon2008}. Furthermore, it is important to note that the hydrodynamics of the plasma and neutral atoms can be quite different. For example the plasma created by a laser or electrical discharge can create shock or sound waves that propagate faster than the plasma expansion.  Therefore, simultaneous imaging of each structure independently is advantageous. By imaging with two wavelengths, the spatial distribution of the electron and neutral background densities may be separated. Dual wavelength interferometry provides a quantitative alternative, but is often not spatially resolved and requires an external reference with interferometric stability \cite{Weber1997}. In this work, we apply recent advances in computational imaging toward improving plasma imaging techniques. 

Coherent diffractive imaging (CDI) is a form of computational imaging that simultaneously retrieves the phase and amplitude of diffracting structures \cite{Fienup1978,Fienup1982c,Elser2003b}. In all CDI methods radiation, across the EM spectrum, probes a sample and diffraction is collected with a pixel detector \cite{Seaberg2011a,Porter2017a,Miao1999}. Propagation of the diffracted radiation is known and thus a phase problem is solved in order to reconstruct an image of the sample \cite{Thibault2008b}. While many forms of CDI exist, ptychography is a particularly robust implementation that allows simultaneous phase and amplitude contrast imaging of extended objects \cite{Rodenburg2007,Thibault2009c}. 

Ptychography involves probing an object with a coherent illumination function over many partially overlapped regions of the specimen.  The redundancy of information from the partially overlapped regions yields a highly constrained phase problem. This allows ptychographic phase retrieval algorithms to produce reliable, high fidelity, noise-robust reconstructions \cite{Bunk2008}. It has been shown that mixed states can be deconvolved, and partial decoherence can be compensated for in advanced reconstruction algorithms, such as ptychographic information multiplexing (PIM) \cite{Thibault2008b,Batey2014,Thibault2013a}.  PIM extended ptychography to allow for simultaneous imaging with multiple wavelength probe illuminations without significant degradation to reconstruction quality as measured by resolution and signal to noise ratio (SNR) \cite{Wei2019}. While these recent developments in advanced ptychographic phase retrieval algorithms are promising for plasma imaging applications, they require transverse scanning of the object thus preventing the possibility of imaging transient, non-reproducible objects. Single-shot ptychography (SSP) overcomes this scanning requirement by breaking up a coherent illumination into smaller beamlets which simultaneously probe the object \cite{Sidorenko2016e,Chen2018,Wengrowicz2019a,Sidorenko2017}. The diffraction patterns from each beamlet are then collected on a single-detector where they are computationally separated to form a data-set similar to that taken in scanning ptychography. Scanning ptychographic phase-retrieval algorithms are used to reconstruct the object. SSP allows for time-resolved ptychographic imaging, making it an applicable imaging technique to probe plasma dynamics.

In this paper, we propose a novel imaging technique in which SSP is performed simultaneously with probes of multiple wavelength. We call this method wavelength-multiplexed single-shot ptychography (WM-SSP). In the methods section, improvements to SSP and the development of WM-SSP are presented. The results section shows a number of simulated and real experiments that convey the SSP improvements and demonstrate WM-SSP applied to both wavelength dependent and non-wavelength dependent objects. In the WM-SSP Plasma Imaging Simulation section a simulated experiment shows how WM-SSP can be applied to calculate the electron and neutral density structure in a plasma. 

\section{Methods}

\begin{figure}[htbp]
\centering
\includegraphics[width=\linewidth]{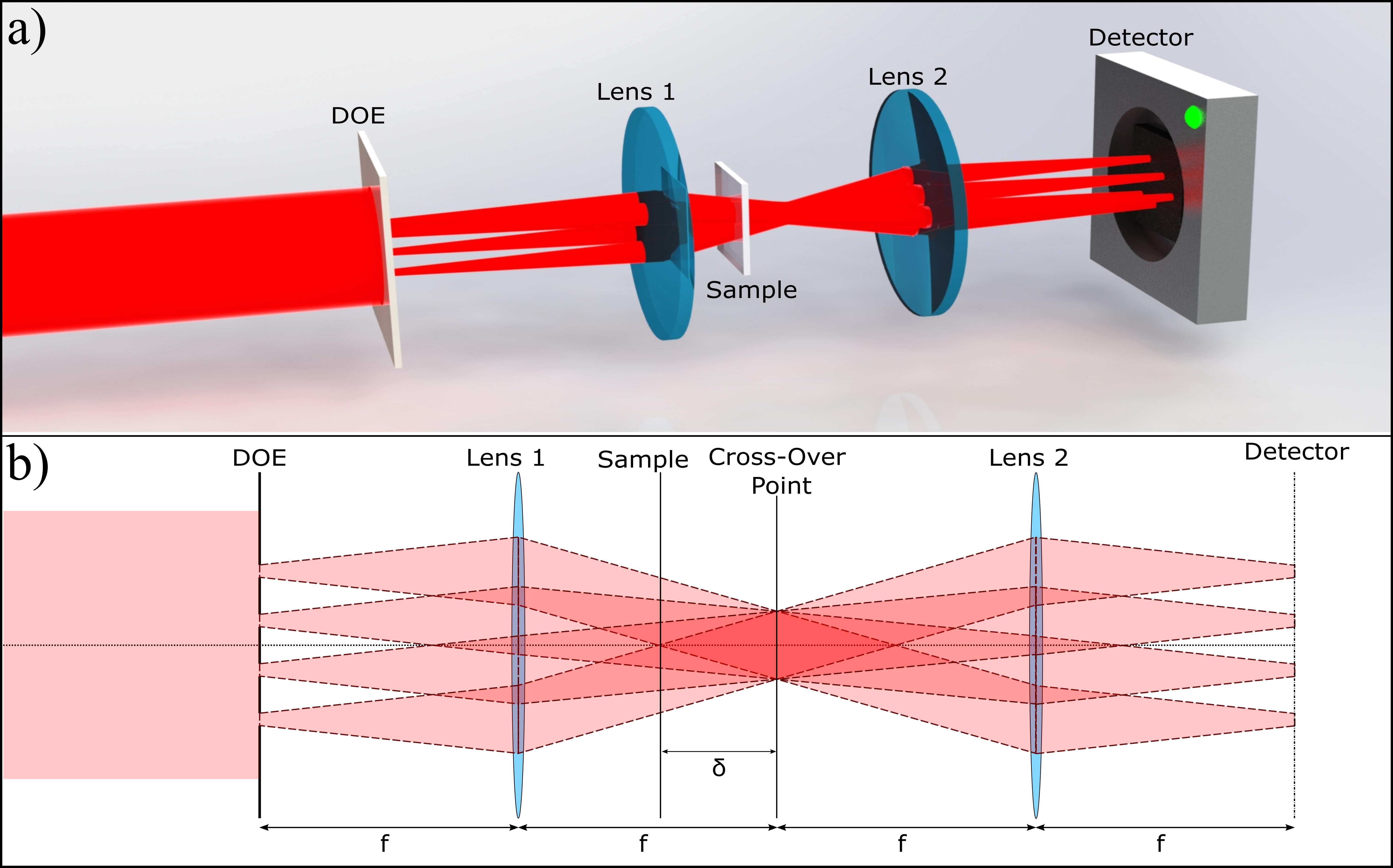}
\caption{In a) a 3D schematic shows the experimental setup with a red illumination. In b) a detailed diagram of the optical setup is shown.  In this optical setup a 4f imaging system is created using two lenses which images the diffractive optical element (DOE) to the detector. The object is placed some distance, $\delta$, away from the crossover point of the beamlets ($\delta$ can be varied to the desired level of probe overlap and diffraction pattern diversity).}
\label{SSPDiagram}
\end{figure}

Before presenting our improvements to single-shot ptychography, we discuss the previously developed method \cite{Sidorenko2016e}. The SSP setup, shown in Figure \ref{SSPDiagram}, consists of a 4f imaging system whereby a diffractive optical element (DOE) is imaged to a detector (camera). The DOE, typically an array of pinholes, breaks up the incident beam into smaller beamlets. These beamlets are collimated and overlap at a point referred to as the crossover point. An object is placed some known distance, $\delta$, away from the crossover point. Beamlets diffract off the object and diffraction pattern intensities are collected around each pinhole image. By computationally segmenting the detector a ptychographic data set is collected in a single-shot. Beamlet locations on the object, analogous to scan positions in scanning ptychography, are calculated using the pinhole image on the detector, the focal length of the second lens, and $\delta$. The diffraction data and calculated probe locations are fed into a conventional scanning ptychographic phase retrieval algorithm, such as the extended  ptychographical  iterative  engine or ePIE algorithm, which simultaneously reconstructs quantitative phase and amplitude images of the object and probe \cite{Gerchberg1971,Maiden2009}.

\subsection{Improvements to Single-Shot Ptychography}

The development of SSP represented a major advance in single-shot CDI. While SSP creates a highly constrained phase problem for a monochromatic probe, information-multiplexing adds complexity that necessitates more sophisticated methods. Here we present a number of improvements to SSP that provide consistently high-fidelity reconstructions of the object and probe. These improvements facilitate information multiplexing in SSP via probes of different wavelength. 


In SSP, diffraction pattern quality has a profound impact on the fidelity and resolution of the reconstructions. Diffraction pattern quality degrades because of factors such as aberrations, back reflections, and SNR. We have designed our SSP microscope with careful consideration of the data collection methods. To mitigate back reflections and spherical aberration we used high quality, achromatic, AR coated lenses as well as a windowless camera. Additional reflections, such as reflections between the object and detector, were removed using tilted ND filters. All of the experimental results in this paper were reconstructed from data taken with a high dynamic range (HDR) algorithm to improve fidelity. The trade off between HDR and single-pulse SSP is a slight reduction in fidelity \cite{Wengrowicz2019a}.


To date many SSP setups use a rectangular DOE, i.e., one that consists of pinholes arranged in a rectangular grid \cite{Sidorenko2016e,Chen2018,Wengrowicz2019a}. In SSP, the DOE pattern is analogous to the scanning pattern in scanning ptychography. It has been shown that a Fermat spiral scanning pattern in ptychography increases the reconstruction fidelity by preventing regularly repeating artifacts and creating more uniformly overlapped probes on the object \cite{Huang2014b}. Furthermore, it is known that the Fermat Spiral offers significantly better performance under conditions with imperfections in the data. Data imperfections are unavoidable in SSP, thus using a DOE with pinholes arranged in a Fermat spiral should be more effective than a rectangular grid DOE \cite{Goldberger2020a}. For our Fermat spiral DOE we considered three design parameters: pinhole diameter, number of pinholes, and pinhole spacing. Here we considered identical pinholes, although theoretically pinholes size and shape variation could be utilized to optimize systematic performance for specific imaging needs. Pinhole diameter was determined from the oversampling condition, given by $\sigma = \frac{\lambda f}{dX D}$ where $dX$ is the pixel size, $D$ is the probe diameter, $\lambda$ is the probe wavelength, and $f$ is the last lens focal length. For our DOE design, we used a pinhole diameter of 55 $\mu$m to give us an oversampling of 8.2. The number of pinholes determines the theoretical lateral resolution given by $Resolution = \frac{\lambda f}{dX N_{px}}$, where $N_{px}$ is the number of pixels in the chopped-out diffraction pattern. Decreasing the number of pinholes increases the chopped-out diffraction pattern size, thus increasing the reconstruction resolution. Therefore, a DOE with only a few pinholes seems favorable, however this must be balanced by the need for redundancy from overlapping probes in a ptychographic data set. Due to these competing effects, we empirically determined that a DOE with 40 pinholes provides a good compromise between resolution and data redundancy for our experimental setup. Finally, pinhole spacing was set such that diffraction signal filled the camera, which minimizes resolution and cross-talk, the coherent superposition of diffraction from adjacent pinholes on the detector. Cross-talk lowers the reconstruction fidelity by introducing mixing of diffraction patterns that is not accounted for in the reconstruction algorithm.


\begin{figure}[htbp]
\centering
\includegraphics[width=\linewidth]{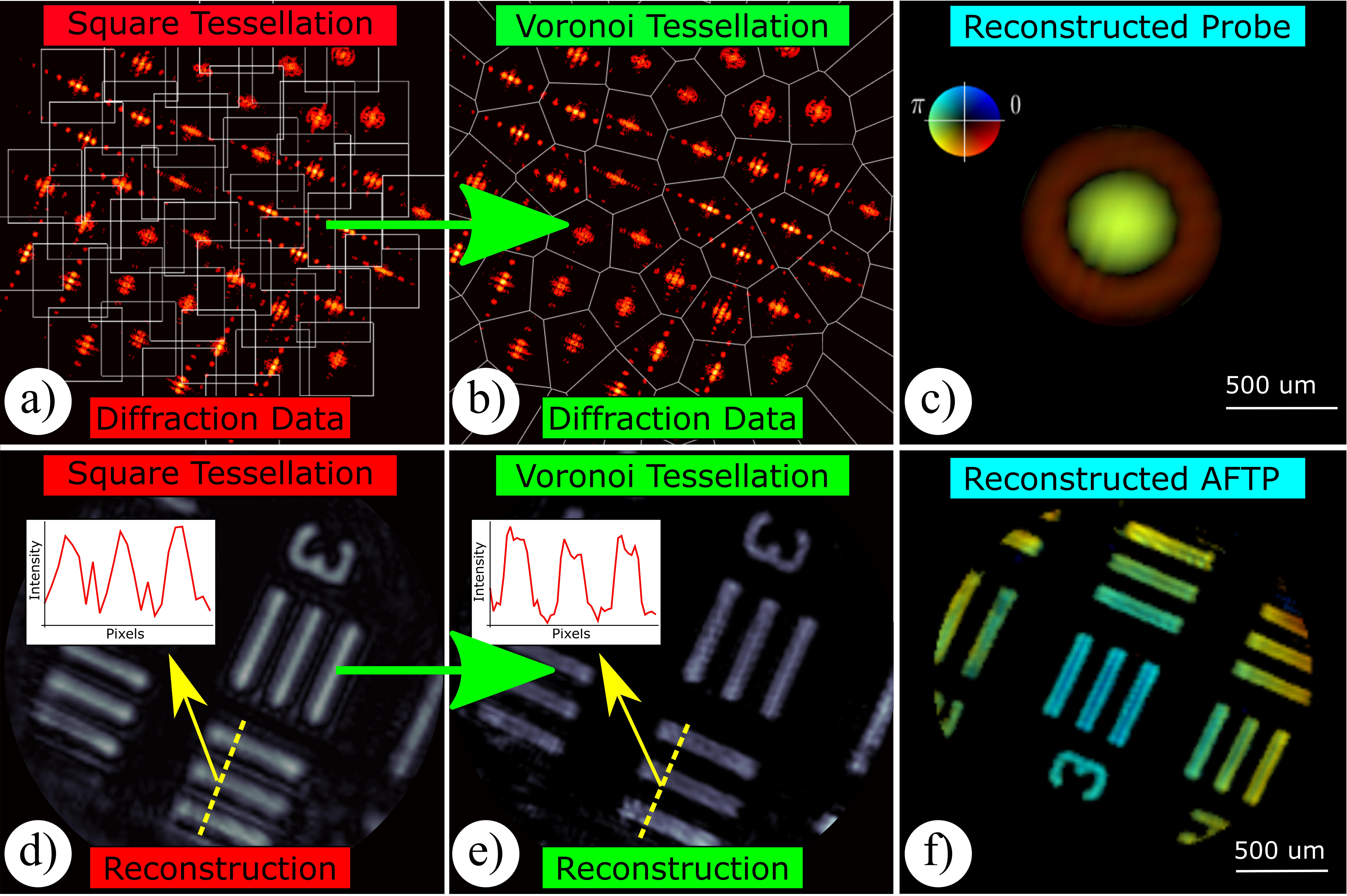}
\caption{Images a) and b) show the square and Voronoi tessellations of diffraction data collected from a SSP system with a Fermat Spiral DOE. The amplitude of the reconstructions from the square and Voronoi tessellated Air Force Test Pattern (AFTP) diffraction data are shown in d) and e). Line-outs from the reconstructions are shown in d) and e), showing that the Voronoi tessellation produces higher fidelity images by removing artifacts and flattening the amplitude in the AFTP bars. Images c) and f) show the complex reconstructed probe and AFTP from a reconstruction with all improvements outlined in the methods section of this paper. The brightness of these reconstructions represent the amplitude and the color represents the phase. The linear phase was removed from the AFTP image caused from tilt of the object.}
\label{VTessellation}
\end{figure}

We process the data by tessellating the detector and segmenting (computationally "chopping out") each beamlet's diffraction pattern. With a rectangular DOE, a square tessellation leads to full use of detector space and no double counting (used in multiple chopped out grids) of diffraction data. Tessellating a Fermat Spiral DOE is more complicated as a square tessellation will lead to unused detector space and/or double counted diffraction data, as illustrated in image a) of figure \ref{VTessellation} . To address this complication, we use a Voronoi tessellation. For a given set of points, $p_{j}$, (the centers of each pinhole) the corresponding Voronoi cells are defined as the set of pixels whose distance from a given point in  $p_{j}$ is equal to or less than the distance from any other point in $p_{j}$. The Voronoi tessellation leads to full utilization of the detector and gives preferential treatment to diffraction closest to each pinhole, shown in image b) of figure \ref{VTessellation}.

\begin{figure}[htbp]
\centering
\includegraphics[width=\linewidth]{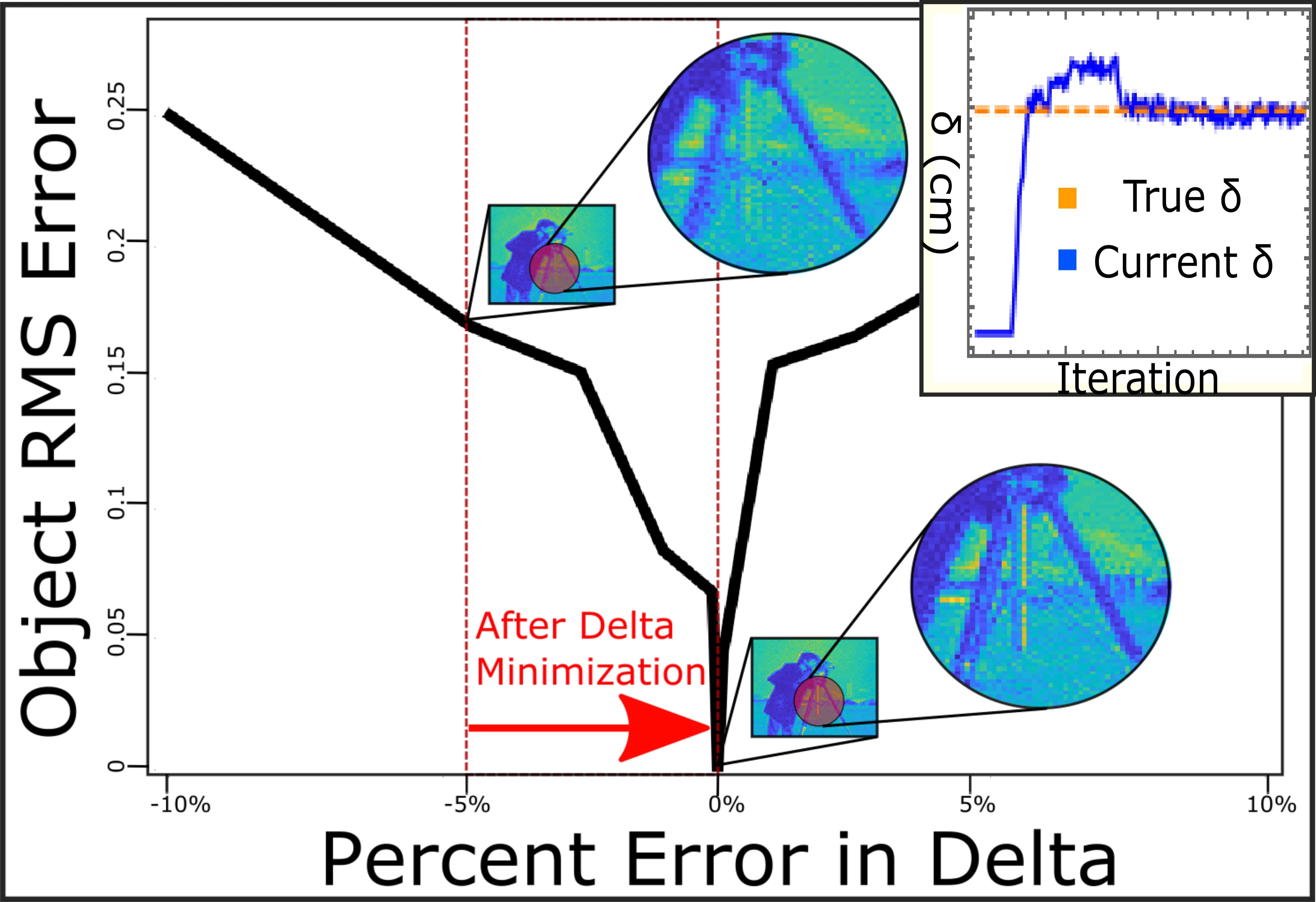}
\caption{Graph of Object RMS Error as a function of the Percent Error in $\delta$. This plot was made from doing reconstructions from the same simulated data set over a range of $\delta$'s. Example reconstructions of the cameraman object are shown from a reconstruction at 5\% error in $\delta$ and 0\% error in $\delta$. The upper-right corner shows the corrected $\delta$ as a function of iteration number from a reconstruction from simulated data. The $\delta$ correction algorithm started on iteration 25 (with a 5\% error in $\delta$) and by iteration 80, it had found the correct $\delta$.}
\label{deltaMinimization}
\end{figure}

Finally, we found that SSP is extremely sensitive to accurate \textit{a priori} knowledge of the distance between the object and the crossover points of the beamlets ($\delta$ as shown in figure 1) (more details in the results section). $\delta$  can be measured accurately using translation stages oriented axially, however this relies on an accurate knowledge of the crossover point's axial location in free space. This works for tangible objects, but is not always possible for transient objects. Therefore, we have developed a phase retrieval algorithm that self corrects for inaccurate $\delta$ information. Since $\delta$ globally scales the beamlets' position on the object, our $\delta$ correction algorithm is based on \cite{Guizar-Sicairos2008}. In our $\delta$ correction algorithm, we start by defining the following per iteration error metric:

\begin{equation}
    \epsilon =  \frac{1}{M N J} \sum_{f_x,f_y,j} [|\Tilde{\psi}_{j}(f_x,f_y)| - I_{j}(f_x,f_y)]^2
\label{errMet}    
\end{equation}

where $\epsilon$ is the error, $M$ and $N$ are the number of pixels in the $X$ and $Y$ directions respectively, $J$ is the number of beamlets, $f_x$ and $f_y$ are the discrete spatial frequency coordinates in the detector plane, $j$ is the beamlet index, $\Tilde{\psi}_j$ is the Fourier transformed exit surface wave of the $j^{th}$ beamlet, and $I_j$ is the measured diffraction pattern intensities from the $j^{th}$ beamlet. Our algorithm solves for $\delta$ by finding the gradient of this error metric with respect to $\alpha$ where $\alpha = \frac{\delta}{f}$, as:

\begin{equation}
    \frac{d \epsilon}{d \alpha} = \frac{1}{M N J} \sum_{f_x,f_y,j} [|\Tilde{\psi}_{j}| - I_{j}]^2 Re{(\Tilde{\psi}_{j}^* \frac{d \psi}{d \alpha}}) 
\label{derOfErrMet}
\end{equation}

where $^*$ represents the conjugate. Here $\frac{d \psi}{d \alpha}$ is defined as:

\begin{equation}
\begin{split}
    \frac{d \psi}{d \alpha} = -2 \pi i  \mathcal{F} ( \mathcal{F}^{-1} ( \Tilde{O}(f_x,f_y) (f_x X_j+f_y Y_j) e^{-2\pi i(f_x X_j+f_y Y_j)})P(x,y))
\end{split}
\label{derOfPsi}
\end{equation}

where $\mathcal{F}$ and $\mathcal{F}^{-1}$ are the 2D-discrete Fourier transform and inverse Fourier transform operators, respectively, $\Tilde{P}$ is the Fourier transformed probe, $X_j$ and $Y_j$ are the beamlet positions, x and y are the discrete object space coordinates, and $O$ is the object. The minimization is then implemented using Newton’s method, and after every iteration of the phase retrieval algorithm delta, is updated in the following way:

\begin{equation}
     \delta_{n+1} = (\alpha_{n} + \beta \frac{d \epsilon}{d \alpha}) f
\label{newtMeth}
\end{equation}

 where $\beta$ is a step-size parameter that determines how much $\delta$ changes in each iteration. Use of this algorithm relaxes the required accuracy of the \textit{a priori} $\delta$ information. Through simulated experiments we found that the algorithm retrieved the simulated $\delta$, when the initial $\delta$ was with 5\% of the simulated $\delta$. The fidelity increase from a reconstruction with 5\% error in $\delta$ and 0\% error in $\delta$ is shown in figure \ref{deltaMinimization}. 

\subsection{Development of wavelength multiplexed single-shot ptychography (WM-SSP)}

In WM-SSP, SSP is performed simultaneously with probes of multiple wavelengths. Instead of using a single-wavelength phase retrieval algorithm like ePIE, an information multiplexed algorithm, PIM, is used to reconstruct objects for each wavelength \cite{Batey2014}. To reinforce WM-SSP, we developed a novel probe constraint specific to the SSP geometry.

In our SSP geometry, pinholes arranged in a Fermat spiral pattern are used to break up the input illumination into multiple beamlets. The pinholes are small relative to the width of the input beam, such that they effectively act as a spatial filter for the beam. All beamlets for a given wavelength are thus the same and do not need to be treated as different in the algorithm (assuming uniform illumination of the DOE). Probes of different wavelengths, however diffract at different spatial rates. The width of the probe (defined by the zeroth order Jinc) is given by:

\begin{equation}
    W_{probe}= 1.22 \frac{\lambda f}{W_{pinhole}}
\label{wProbe}
\end{equation}

where $w_{probe}$ is the width of the probe, $W_{pinhole}$ is the width of the pinhole. The pixel size in object space, the basic unit of the detector's pixelated grid propagated to the object, is defined as:

\begin{equation}
    dX_{object}= \frac{\lambda f}{W_{detector}}
\label{ObjPx}
\end{equation}

where $dX_{object}$ is the object space pixel size, and $W_{detector}$ is the size of a chopped out diffraction pattern. Both the width of the probes and the object space pixel size scale proportional to $\lambda$. Therefore, in pixel space, probes of different wavelengths will be exactly the same, since the factors of $\lambda$ cancel. In every iteration of the phase retrieval algorithm, we can enforce this constraint by averaging the updated probes for each wavelength. The spectral weights i.e., power contained in each probe wavelength, can still be calculated in each iteration separately, but the spatial profile is constrained to be the same in pixel space for each wavelength. This probe constraint allows the phase retrieval algorithm to give high fidelity reconstructions for each wavelength. 

\section{Results}

\subsection{Demonstration of Improvements to Single-Shot Ptychography}

In the previous section, we presented the use of a Fermat spiral DOE, a Voronoi tessellation to segment the detector, and a $\delta$ minimization algorithm to correct for inaccurate \textit{a priori} knowledge of $\delta$. Here we present simulated and experimental demonstrations that illustrates how these additions to the SSP technique improves the method and reconstruction fidelity.

The use of a Fermat Spiral probing pattern in scanning ptychography has been well studied and is known to increase the fidelity of reconstructions. As explained in the Methods section, this implies that using a DOE with a Fermat Spiral pinhole pattern will increase the reconstruction image quality. To show that a Voronoi tessellation, when using a Fermat Spiral DOE, increases reconstruction fidelity, we conducted an experiment using a real world setup. The 4f-imaging system in the setup consisted of two 5 cm air-spaced achromatic lenses, Thorlabs part number ACA254-050-A, with high quality AR coatings at the probing wavelengths. The probing laser was a 532 nm CW-diode laser, which was spatially filtered, expanded and collimated to create relatively uniform illumination on the DOE. The DOE, (a custom made photo-lithography mask: HTA Photomask) was made-up of 40 pinholes each with a 55 $\mu$m diameter arranged in a Fermat spiral. The detector (Thorlabs 8051M-USB) was an 8 mega-pixel camera with the wedge and face-plate removed. The camera sensor array has 3296x2472 square pixels with a pixel size, $dX$ of 5.5 $\mu$m. The DOE under-filled the sensor so that the diffraction filled around 50\% of the total detector space. 

To compare square and Voronoi tessellations, an Air Force Test Pattern (AFTP) was imaged at $\delta = 0.9 cm$. Holding everything else constant, the collected diffraction intensities were segmented using both square and Voronoi tessellations. Since the Fermat spiral is not regularly spaced, we chose a square tessellation with cells of 300x300 pixels to balance the effects of double-counting diffraction data and not using enough diffraction data. The Voronoi tessellation will create cells of varying sizes that are chopped out and placed on a square cell of 444x444 pixels (padded with zeros). Reconstructions were performed using both tessellations for 1000 iterations, under the same reconstruction parameters. The amplitude of the retrieved images are shown in d) and e) of figure \ref{VTessellation}. The reconstruction from the square tessellation has artifacts, like the lines between the real lines of the AFTP, and low fidelity as seen in the non-uniform amplitude distribution within the bars of the AFTP. The Voronoi tessellated reconstruction removes these artifacts and has significantly better fidelity of the flat top AFTP bars. This can be seen more easily in the line outs shown in d) and e) of figure \ref{VTessellation}. 

The fidelity of a SSP reconstruction is heavily dependent on accurate\textit{a priori} knowledge of $\delta$. To show this, we performed a simulated experiment where reconstructions were done over a range of $\delta$'s. The simulated data was collected with a detector made up of 2048x2048 square pixels of $dX = 10 \mu m$. The 4f-imaging system was made up of two 10 cm focal length lenses and the simulated probe wavelength was 532 nm. The simulated DOE consisted of 30 circular pinholes with 50 $\mu m$ diameters arranged in a Fermat Spiral spaced to fill the detector. The object imaged was the cameraman placed at a $\delta$ of 0.9 cm outside of the crossover point. The field was propagated between lenses using normalized FFTs and between the crossover point and the object plane using spectrum of plane waves. With the simulated diffraction patterns, reconstructions were done from 10\% below to 10\% above the simulated $\delta$. The root mean square (RMS) error was calculated between the reconstructed object at the simulated $\delta$ and the reconstructed object at every sampled $\delta$. A plot of the RMS error vs Percent Error in $\delta$ is shown in figure \ref{deltaMinimization}. The graph shows a drastic increase in object RMS error for small errors in $\delta$, which is reflected in the decrease in fidelity of the cameraman shown in the reconstructions at 5\% error in $\delta$. The $\delta$ correction algorithm, explained previously, was then tested on the simulated data used for this experiment. Starting with a 5\% error in $\delta$, the correction algorithm was able to converge on the correct $\delta$ after 55 iterations. A graph showing $\delta$ change as a function of reconstruction iteration is shown in the inset of figure \ref{deltaMinimization}. Our simulation shows that the $\delta$ correction algorithm was able to retrieve the correct $\delta$ when the guess was within 5\% of the correct $\delta$.

With all these improvements to SSP, an experimental reconstruction of an AFTP was done using the system previously described for the Voronoi Tessellation experiment. The complex probe and AFTP are shown in c) and f) of figure \ref{VTessellation} respectively. In these reconstructions, the brightness represents the amplitude of the object and the color represents the phase. We removed the linear phase on the object which was caused from a tilt of the object with respect to the optical axis. The quadratic phase that is present in the object reconstruction is believed to come from the angled probes present in SSP. These images show the high fidelity reconstructions that are consistently retrieved with our SSP microscope. 

\subsection{Experimental Verification of wavelength-multiplexed single-shot ptychography: test pattern}

\begin{figure}[htbp]
\centering
\includegraphics[width=\linewidth]{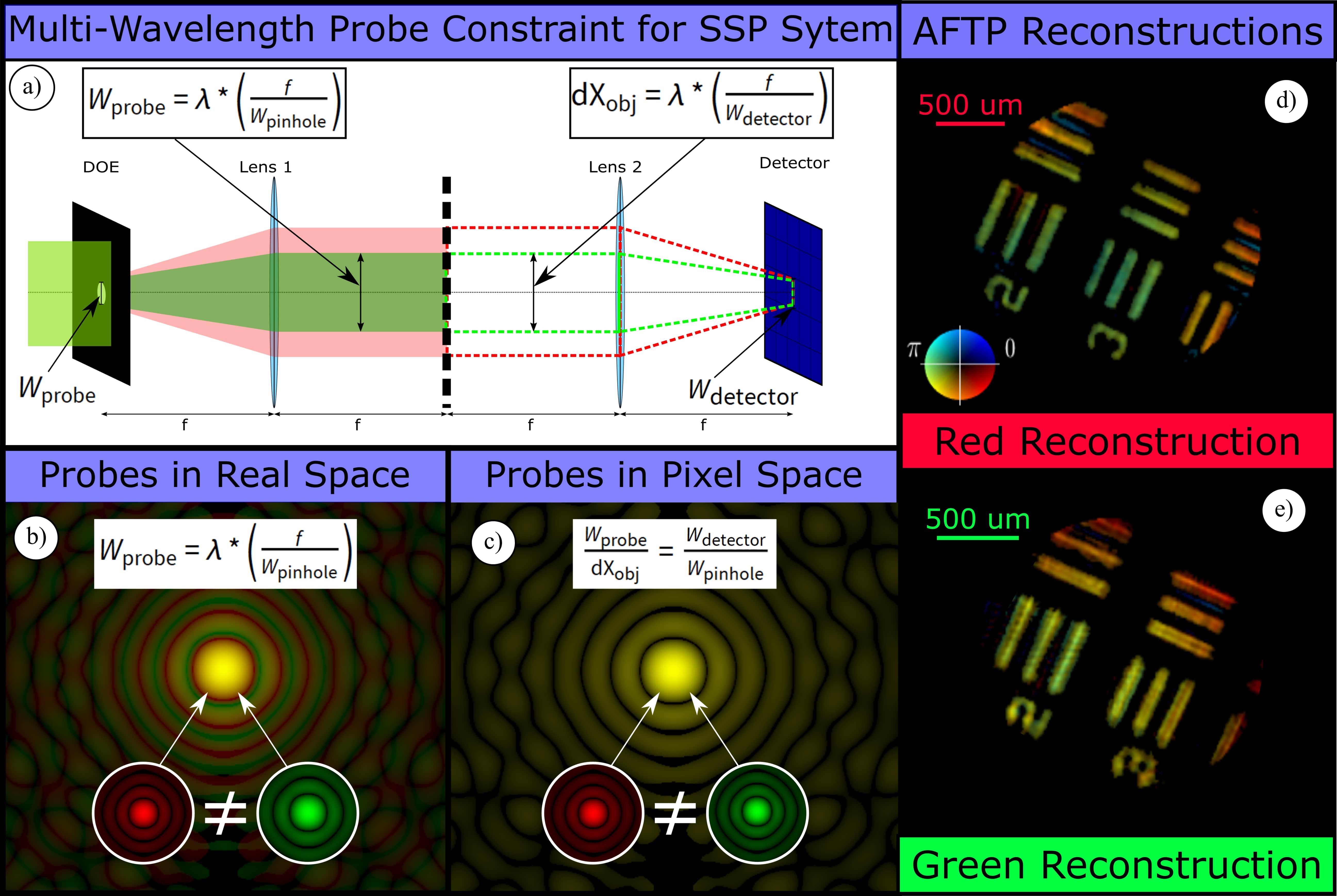}
\caption{The novel probe constraint is pictorially represented in a). The probe width and the object space pixel size vary linearly with the probe wavelength. The probes in real space units are shown in b), where the red and green probes are not overlapped, due to the wavelength dependence of the probe widths. The probes in pixel space are shown in c), where the red and green probes are spatially overlapped due to the wavelength dependencies of the probe width and object space pixel size cancelling. Complex Experimental WM-SSP AFTP reconstructions are shown in d) and e) (the amplitude is represented by the brightness while the phase is represented by the color). The difference in field of view is due to the wavelength dependence of the object space pixel size.}
\label{MultiWavelengthDevFig}
\end{figure}

As an initial test of WM-SSP, we reconstructed an object with no wavelength dependence. The chosen object was an AFTP deposited on glass, which has a negligible amount of spatial dispersion from the beamlets incident on the glass at an angle. The data was collected with 532 nm and 633 nm CW-lasers simultaneously illuminating the experimental SSP system previously described. WM-SSP reconstructions were performed using an algorithm based on PIM with the novel probe constraint. Ghost modes were also used to produce higher fidelity reconstructions. A ghost mode is an additional mode for every wavelength that is summed incoherently at the detector. Every iteration of the phase retrieval algorithm, the ghost mode reconstructed object is flipped across the x-y axis and laterally shifted \cite{rana}. The reconstructed complex AFTP for the 532 nm and 633 nm probes are shown in figure \ref{MultiWavelengthDevFig}. The different field-of-view for the reconstructions is expected due to the difference in probe and object-space pixel sizes shown in equations \ref{wProbe} and \ref{ObjPx}. 

\subsection{Experimental verification of wavelength-multiplexed single-shot ptychography: wavelength-dependent object}

\begin{figure*}[htbp]
\centering
\includegraphics[width=\linewidth]{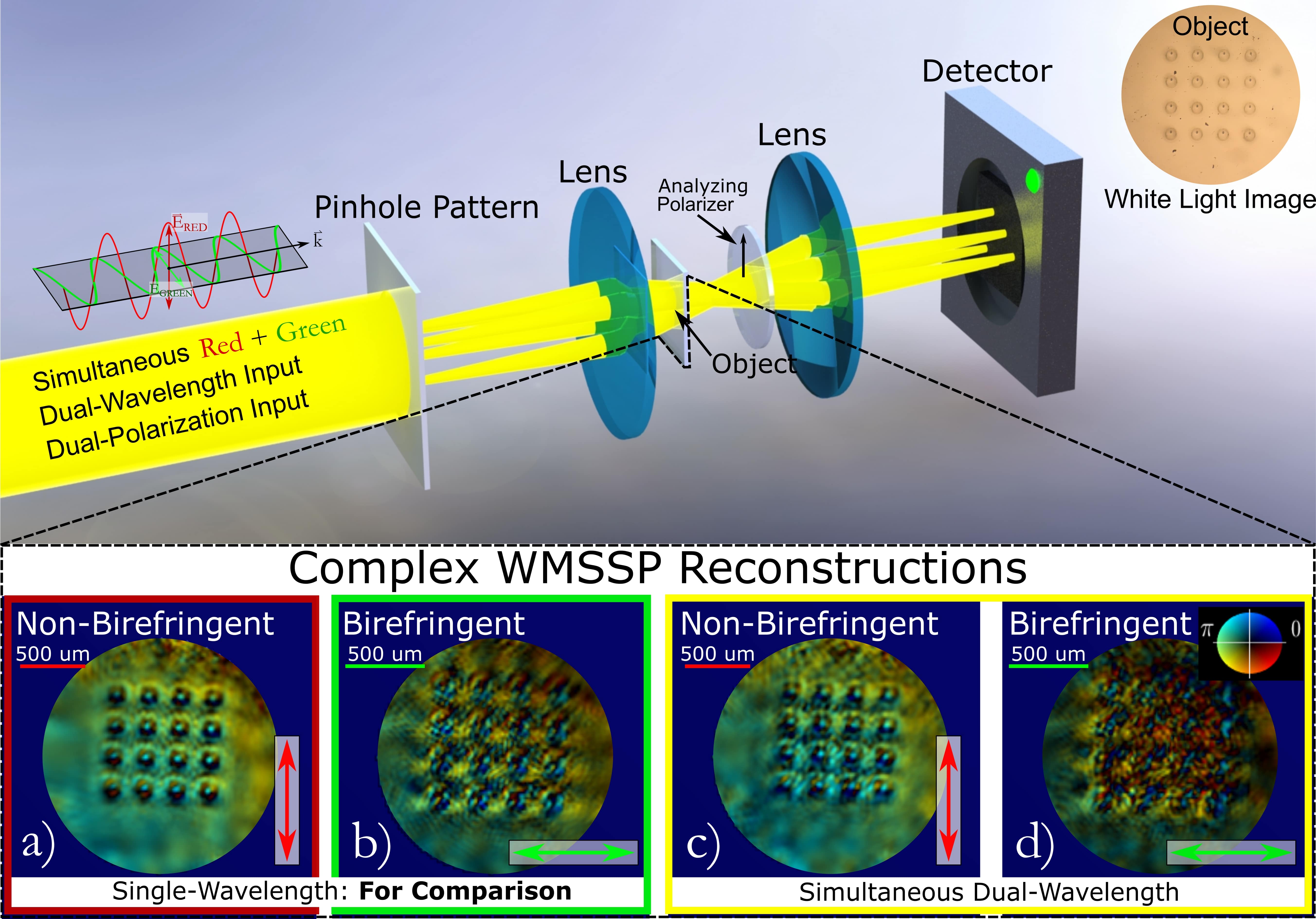}
\caption{The WM-SSP setup for imaging the birefringent laser machined sample is pictorially represented. The red probe illumination is vertically polarized and the green probe illumination is horizontally polarized. By aligning an analyzing polarizer after the birefringent object, the red probe images the non-birefringent part of the sample while the green probe images the birefringent part of the sample. The complex object reconstructions from the single-wavelength data are shown in a) and b) for comparison against the complex reconstruction from the multi-wavelength data shown in c) and d) (intensity represents the amplitude while the color represents the phase). A white light image of the laser-written pattern is shown to the left of the reconstructions for reference. Comparing c) to a) and d) to b) shows that WM-SSP reproduces high fidelity reconstructions of wavelength-dependent objects.}
\label{BirefringentMultiWavelength}
\end{figure*}

We experimentally verified WM-SSP by reconstructing a wavelength-dependent object. The setup is shown in figure \ref{BirefringentMultiWavelength} and is the same as previously described with the addition of an analyzing polarizer. The wavelength-dependent object was made by exploiting the resulting stress-induced birefringence from femtosecond laser exposure of transparent substrates \cite{Champion2013}. The sample imaged is composed of a 4x4 grid of cylinders (imaged along the axis so they appear to look like circles) inscribed in a soda-lime substrate using a femtosecond laser. (Details about the fabrication methods for these structures can be found in \cite{McMillen2014,McMillen2015}). In the case of soda-lime and for the laser exposure conditions used here, the cylinders produce an isotropic stress distribution. Specifically, each cylinder has a diameter of 100 microns and consists itself of a stack of 128 laser-written layers. By simultaneously illuminating the sample with a horizontally polarized 532 nm probe and vertically polarized 633 nm probe and placing an analyzing polarizer after the object, birefringent and non-birefringent objects are imaged separately. The analyzing polarizer was aligned to the vertical polarization such that the 633 nm probe imaged the non-birefringent parts of the sample while the 532 nm probe imaged the birefringent object. Data was taken with both beams simultaneously illuminating the object and each beam individually imaging the object. We reconstructed the single-wavelength data sets, each running 5000 iterations of a standard ePIE algorithm. These reconstructions serve both as a base to compare the WM-SSP reconstructions against but also as object guesses for the WM-SSP retrieval algorithm to start with. Reconstructions of the dual-wavelength data set was done, running 10,000 iterations. The novel probe constraint and ghost modes were used in this reconstruction. The complex single-wavelength reconstructions and multi-wavelength reconstructed objects are shown in figure \ref{BirefringentMultiWavelength}. The 633 nm reconstructions, non-birefringent part of the objects, both show a 4X4 array of circles with dots in the middle. The 532 nm reconstructions, birefringent objects, are more complicated but share the same diagonal features going through the 4X4 array of laser machined circles (cylinders viewed from the top). Since the single-wavelength and multi-wavelength reconstructions compare favorably we are confident that WM-SSP works for wavelength dependant objects. This experiment also provides a novel method for both space and time-resolved quantitative analysis of stress-strain dynamics of laser machining process. This technique has the capability of providing laser machinist valuable information about the stress-induced birefringence they are creating in their machining processes.

\section{WM-SSP Plasma Imaging Simulation}

One exciting application of WM-SSP is plasma dynamics analysis. Plasma are created by ionizing gas, creating a mix of neutral unionized atoms, ions, and electrons. The neutral background atoms exhibit different hydrodynamics than the electronic structures and thus in observing the dynamics involved in plasma formation it is favorable to image each species separately. This is difficult due to their spatial overlap, non-reproducible dynamics, and fast time-scales. One way to observing the electron's and neutral's behavior independently is using dual-wavelength interferometry. Dual-wavelength interferometry utilizes the electrons' wavelength dependent index of refraction to separate the contribution to the phase from the electrons and neutrals \cite{Weber1997}. The phase shift from the electrons, relative to vacuum, is given by:

\begin{equation}
    \Delta \phi_e = - \frac{e^2 \lambda_0 \rho_e \Delta Z_e}{2 \pi \epsilon_0 m_e c^2}
\label{ePhase}
\end{equation}

where $e$ is the fundamental charge, $\epsilon_0$ is the vacuum permittivity, $m_e$ is the mass of an electron, $c$ is the speed of light, $\lambda_0$ is the probe wavelength, $\rho_e$is the electronic density, and $\Delta Z_e$ is the plasma thickness (assuming a uniform plasma in the axial direction). The phase shift from the neutral background, relative to vacuum, is given by:

\begin{equation}
    \Delta \phi_N = - \frac{4 \pi (n_0-1) \Delta Z_N}{\lambda_0}  (1 - \frac{\rho_N}{\rho_0})
\label{nPhase}
\end{equation}

where $n_0$ is the index of refraction of the neutral background, $\rho_N$ is the density of the neutral background, $\rho_0$ is the density of the gas the plasma is formed inside, and and $\Delta Z_N$ is the thickness of the uniformly distributed neutrals \cite{Weber1997}. Dual-wavelength interferometry experimentally determines the phase shifts for two wavelengths and uses equations \ref{ePhase} and \ref{nPhase} to setup a system of equations to solve for $\rho_e$ and $\rho_N$. Generally dual-wavelength interferometry (DWI) is not spatially resolved. Spatially (Two-dimensional) and temporally resolved versions of DWI, and generally most plasma imaging methods, are based on pump-probe setups which preclude them from studying non-reproducible plasma dynamics \cite{Hayasaki2017,Aragon2008}. Since WM-SSP numerically solves for the object phase at each wavelength, the same system of equations can be setup and solved to calculate the spatially- and temporally- dependent $\rho_e$ and $\rho_N$ from a single WM-SSP data set. To show the potential of this technique, we performed an experimental simulation.

\begin{figure}[htbp]
\centering
\includegraphics[width=\linewidth]{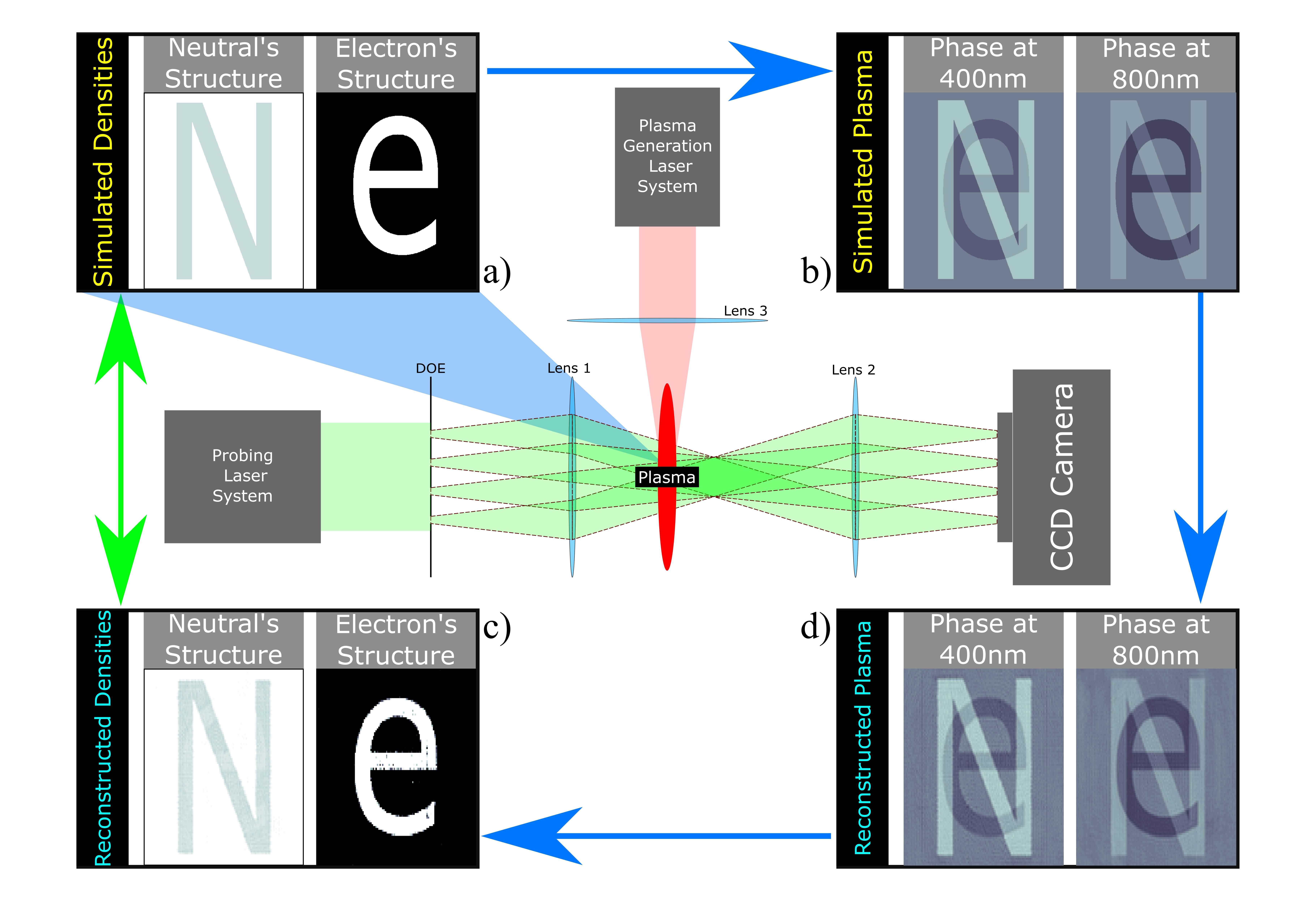}
\caption{The simulated electron and neutral structure is shown in a). This plasma was probed by 400 nm and 800 nm coherent illumination in simulation, shown in b). A simulated WM-SSP diffraction data set was collected and reconstructed. The reconstructions are shown in c). The plasma's electron and neutral structure was calculated, shown in d), from the reconstructed object.}
\label{PlasmaDeconFigure}
\end{figure}

The simulated WM-SSP setup included two 100 cm-focal length lenses, a detector with 2048x2048 square pixels of $dX = 50 \mu m$, and a DOE with 20 circular pinholes of $130 \mu m$ diameter arranged in a Fermat spiral. The plasma was simulated with $\rho_e = 2.65 * 10^{24}$ electrons/m$^3$, $n_0 = 1.0005$, $\rho_N = 2.386*10^{25}$ atoms/m$^3$, $\rho_0 = 2.65*10^{25}$ atoms/m$^3$, $\Delta Z_e = 10^-4$ m$^3$, and $\Delta Z_N = 10^-3$ m$^3$. These simulation parameters were chosen to represent a plasma being formed in air at STP with an ionization fraction of $10\%$.  The electron's spatial structure was made to resemble an 'e' (everything inside the e was uniformly distributed electrons and everything outside did not contributing to the electron phase shift). The neutral's structure was made to resemble an 'N' (everything inside the N was uniformly distributed neutral atoms and everything outside the N was uniformly distributed background atoms). The plasma was simulated with unit amplitude such that the complex transfer function was written as Object$_{plasma} = e^{i\Delta \phi_e (x,y) + i\Delta \phi_N (x,y)}$. The simulated electron structure, neutral structure, and plasma object are shown in figure \ref{PlasmaDeconFigure}. Diffraction data was simulated by placing the 2D plasma object at $\delta = 4$ cm and probing it simultaneously with 400 nm and 800 nm probe illumination. The plasma objects were reconstructed at each probe wavelength and the electron and neutral densities were calculated. The reconstruction went for 10,000 iterations, the simulated probes were used, the object amplitude for each wavelength was enforced to be one everywhere, and no ghost modes were used. The reconstructed plasma and calculated densities are shown in figure \ref{PlasmaDeconFigure}c and figure \ref{PlasmaDeconFigure}d respectively. The simulated experiment shows the high fidelity electron and neutral density structures that can be calculated from a single WM-SSP data set. The development of WM-SSP represents a major step in computationally studying the dynamics involved in plasma formation.

\section{Conclusion}

Through a number of improvements to SSP, we demonstrated wavelength multiplexing in single-shot ptychography. These improvements made SSP a more reliable method of single-shot CDI. Through the development of a novel probe constraint, in addition to the improvements made to SSP, wavelength-multiplexed single-shot ptychography was experimentally realized. By imaging a birefringent object with multiple wavelengths that were orthogonally polarized, we demonstrated our novel computational imaging method by imaging a wavelength dependent object. After reconstructing both wavelengths individually and simultaneously, the reconstructions of each wavelength favorably compared thus verifying that wavelength-multiplex SSP works. WM-SSP was applied to image a simulated plasma composed of both electrons and neutral atoms. Exploiting the electrons' wavelength dependent nature allowed us to calculate the plasma's electron and neutral density structures from a WM-SSP reconstruction. This imaging method is ready to be applied to imaging transient phenomena such as dynamically evolving plasmas. Wavelength-multiplexed SSP allows for quantitative (resolved in both 2D space and time) imaging of both the electron densities and the neutral densities within a single plasma event. Furthermore, this method has the capability of probing quantitatively for fundamental proprieties of plasma, making WM-SSP a powerful novel plasma imaging metrology.

\section*{Funding}
The authors gratefully acknowledge funding from the Air Force through AFOSR FA9550-18-1-0089 and Los Alamos National Laboratory through contract number 501188. 

\section*{Disclosures}
The authors declare no conflicts of interest.

\bibliographystyle{ieeetr}
\bibliography{biblioWMSSP.bib}






\end{document}